\documentclass[prb,twocolumn,tbtags,superscriptaddress,floatfix,showpacs]{revtex4}
\usepackage{amssymb}
\usepackage{graphicx,amsmath}
\usepackage{bm}
\usepackage{times}

\begin{document}

\title{Resonance at the Rabi frequency in a superconducting flux qubit}

 \keywords      {Superconducting qubits,
Rabi frequency, amplification, attenuation, waveguide,
transmission line.}

\begin{abstract}
 We analyze a system composed of a superconducting flux qubit coupled
to a transmission-line resonator driven by two signals with
frequencies close to the resonator's harmonics. The first strong
signal is used for exciting the system to a high energetic state
while a second weak signal is applied for probing effective
eigenstates of the system. In the framework of doubly dressed
states we showed the possibility of amplification and attenuation
of the probe signal by direct transitions at the Rabi frequency.
We present a brief review of theoretical and experimental works
where a direct resonance at Rabi frequency have been investigated
in superconducting flux qubits. The interaction of the qubit with
photons of two harmonics has prospects to be used as a quantum
amplifier (microwave laser) or an attenuator.
\end{abstract}

\pacs{85.25.Dq,~ 85.25.Cp,~ 85.25.Hv,~ 84.40.Az}

\date{\today }
\author{Ya. S. Greenberg}\email{yakovgreenberg@yahoo.com}
\affiliation{Novosibirsk State Technical University, Novosibirsk,
Russia}
\author{E.~Il'ichev}
\affiliation{Institute of Photonic Technology, Jena,
Germany}
\author{G. Oelsner}
\affiliation{Institute of Photonic Technology, Jena,
Germany}
\author{S. N. Shevchenko}
 \affiliation{B. Verkin
Institute for Low Temperature Physics and Engineering, Kharkov,
Ukraine} \affiliation{V. Karazin Kharkov National University,
Kharkov, Ukraine}


 \maketitle


\section{Introduction}

Superconducting qubits are known as the key elements for the solid
state implementation of quantum computers
\cite{Clarke08,Grajcar05}. Among them superconducting flux qubits
are most promising ones due to their relatively long dephasing
times and robustness against external disturbances. Since their
invention \cite{Mooij99, Orlando99} flux qubits have been
intensively investigated mainly due to their multiple potential
applications, ranging from microwave quantum metamaterials
\cite{Macha13} to ultrahigh-sensitive magnetometers
\cite{Ilichev07}.

A superconducting flux qubit (or, simply, flux qubit) is a
superconducting loop interrupted by three Josephson junctions. Two
junctions are identical while the critical current of the third
junction is smaller by approximately twenty percent. In energy
language such a structure exhibits two energy levels which are
formed by a superposition of quantum states corresponding to
clockwise and counter clockwise directions of a superconducting
current $I_q$ along the loop. The interlevel distance is as
follows:
\begin{equation}\label{1}
    \Delta E=\hbar\sqrt{\varepsilon^2+\Delta^2},
\end{equation}
where $\varepsilon$ is an external parameter which by virtue of
external magnetic flux, $\Phi_X$ controls the gap between ground
and excited states \cite{Wal00}:
\begin{equation}\label{2}
\varepsilon=\frac{2I_q}{\hbar}\left(\Phi_X-\frac{\Phi_0}{2}\right),
\end{equation}
where $\Phi_0=h/2e$ is a flux quantum. The quantity $\Delta$ in
(\ref{1}) is the gap between ground and excited states at the
degeneracy point ($\varepsilon=0$).

  Characteristic energies (in a frequency scale) of superconducting qubits
belong to the GHz  frequency range. Nevertheless quantum
properties of superconducting flux qubits have been successfully
investigated in a MHz frequency range by so called impedance
measuring technique where the flux qubit has been readout with a
low frequency (about hundred MHz) LC tank circuit
\cite{Ilichev02,Greenberg02,Greenberg02a,Grajcar04,Ilichev04}.

From a formal point of view a superconducting qubit is analogous
to a spin-1/2 particle with Hamiltonian
\begin{equation}\label{3}
    H_q=\frac{\hbar\omega_q}{2}\sigma_Z,
\end{equation}
where $\omega_q=\Delta E/\hbar$.

Many experiments have shown that the interaction between a flux
qubit and microwave radiation can be described and interpreted
within the models and approaches well known from quantum optics.
That is why flux qubits are frequently called artificial atoms.
However, there are two significant differences. The first one is
that a flux qubit is a macroscopic device, hence, its interaction
with microwaves is much stronger than that for the atom-photon
interaction. For flux qubit the regime of strong coupling to a
photon field can easily be achieved. For this reason the effects
which are observed in a media with many atoms can be observed for
a single flux qubit interacting with a microwave field
\cite{Astafiev10}. The second difference relates to the form of
the interaction between a flux qubit and a photon field. The
interaction Hamiltonian is as follows \cite{Omel10}:
\begin{equation}\label{H_Int}
    H_{int}=\hbar g_{ } \left(\frac{\varepsilon }{\omega _q }\sigma _Z +
\frac{\Delta }{\omega _q}\sigma _X\right)\left( {a^ +   + a}
\right),
\end{equation}
where $a^+, a$ are creation and annihilation operators for
photons, $g$ is the coupling strength between qubit and photon
field.

Hamiltonian (\ref{H_Int}) contains the "longitudinal"  term $\hbar
g \varepsilon \sigma _Z \left( {a^ + + a} \right)/\omega_q$ the
analogue of which is absent in quantum optics due to antisymmetric
properties of the atom dipole moment. This term gives rise to the
transitions between Rabi levels of the same doublet, the effect
which is not allowed for atoms in photon field. This effect was
first observed long time ago in low frequency NMR experiments
\cite{Mefed77,Mefed78} where a time dependent part of the
polarizing magnetic field $B_0$ played the role of our parameter
$\varepsilon$ in (\ref{H_Int}) (see also section VIIB in
\cite{Greenberg07}).

The first experimental detection of the voltage fluctuations at
the Rabi frequency in low frequency LC circuit coupled to a flux
qubit was observed in \cite{Ilichev03}. This paper gave rise to a
series of theoretical works devoted to the transitions between
Rabi levels in superconducting flux qubit. In  \cite{Greenberg05}
it was shown that in the dissipative two level system it is
possible to obtain persistent Rabi oscillations excited by an
external signal which is in resonance with Rabi splitting. It was
also shown that in a dissipative two level system the regimes of
amplification and attenuation of a probe signal due to transitions
between Rabi levels of the same pair were possible
\cite{Greenberg07, Greenberg08, Hauss08a, Hauss08b}.  In these
papers two- probe spectroscopy was considered with a driving
signal controlling the populations of the Rabi levels. In
Ref.\cite{Greenberg07} the probe signal was considered as
classical wave, while in the works \cite{Hauss08a, Hauss08b} the
probe signal was a photon field. All these theoretical findings
have been confirmed in experiment  \cite{Grajcar08} where a flux
qubit driven by a high-frequency field with frequency $\omega_d$
was coupled through a mutual inductance to a low-frequency tank
circuit with frequency $\omega_T$ much lower than the qubit gap
$\Delta$. In this experiment the amplification and attenuation of
the LC voltage by a flux qubit are clearly due to the transitions
between nearby Rabi levels with the splitting being in resonance
with the LC tank.

With the progress in thin film technology it became possible to
develop a microstrip waveguide with a very high quality factor
\cite{Macha10}. This, in turn, paved the way to many interesting
experiments where one or several qubits were coupled on-chip to a
microwave resonator (see, for example, Ref. \cite{Omel10} and
references therein). In the context of the subject of this review
it is necessary to mention the paper \cite{Oelsner13} where the
first experimental demonstration were given of the amplification
and attenuation of microwaves due to Rabi transitions between the
levels of the same Rabi doublet. In Ref. \cite{Oelsner13} a flux
qubit was placed in the middle of coplanar half-wavelength
waveguide which had two gaps at its ends. For this reason the
waveguide plays for microwaves the same role as a Fabry-Perot
resonator plays for photons in quantum optics. The waveguide was
driven with a strong signal at its third harmonic. A weak probe
signal at the fundamental frequency was detected at the output
where the phase and the amplitude of the transmission factor were
measured. Because the Rabi splitting can be controlled by several
external parameters it can be matched to resonance with the
fundamental frequency resulting in attenuation or amplification of
the probe signal. The direction of this effect depends on the
populations of the Rabi levels which in turn depend on the
external parameters and on the rates of relaxation and dephasing
of the flux qubit. This experiment has been carefully analyzed in
Ref. \cite{Shevchenko14}. Here we give a brief account of this
analysis from the point of the dressed state approach which is
well known in quantum optics (see, for example, Ref.
\cite{Coh-Tan}).

\section{superconducting flux qubit and photon field}

\subsection{Interaction of a superconducting flux qubit with a
strong driving microwave field}

First we consider a superconducting flux qubit which interacts
with a one-mode quantized driving photon field.

\begin{eqnarray}\label{4}
& H_{qb-d} = \frac{{\hbar \omega _q }}{2}\sigma _Z  + \hbar \omega
_d d^ + d + \hbar g_{_d } \frac{\varepsilon }{{\omega _q }}\sigma
_Z \left( {d^ +   + d} \right)+ & \nonumber \\
& + \hbar g_{_d } \frac{\Delta }{{\omega _q }}\sigma _X \left( {d^
+ + d} \right),&
\end{eqnarray}

where $d^+, d$ are creation and annihilation operators for the
field mode with an angular frequency $\omega_d$. The third and
fourth terms represent the coupling between the flux qubit and the
driving field with $g_d$ being the coupling strength.

\begin{figure}
  \includegraphics[height=.4\textheight, angle=-90]{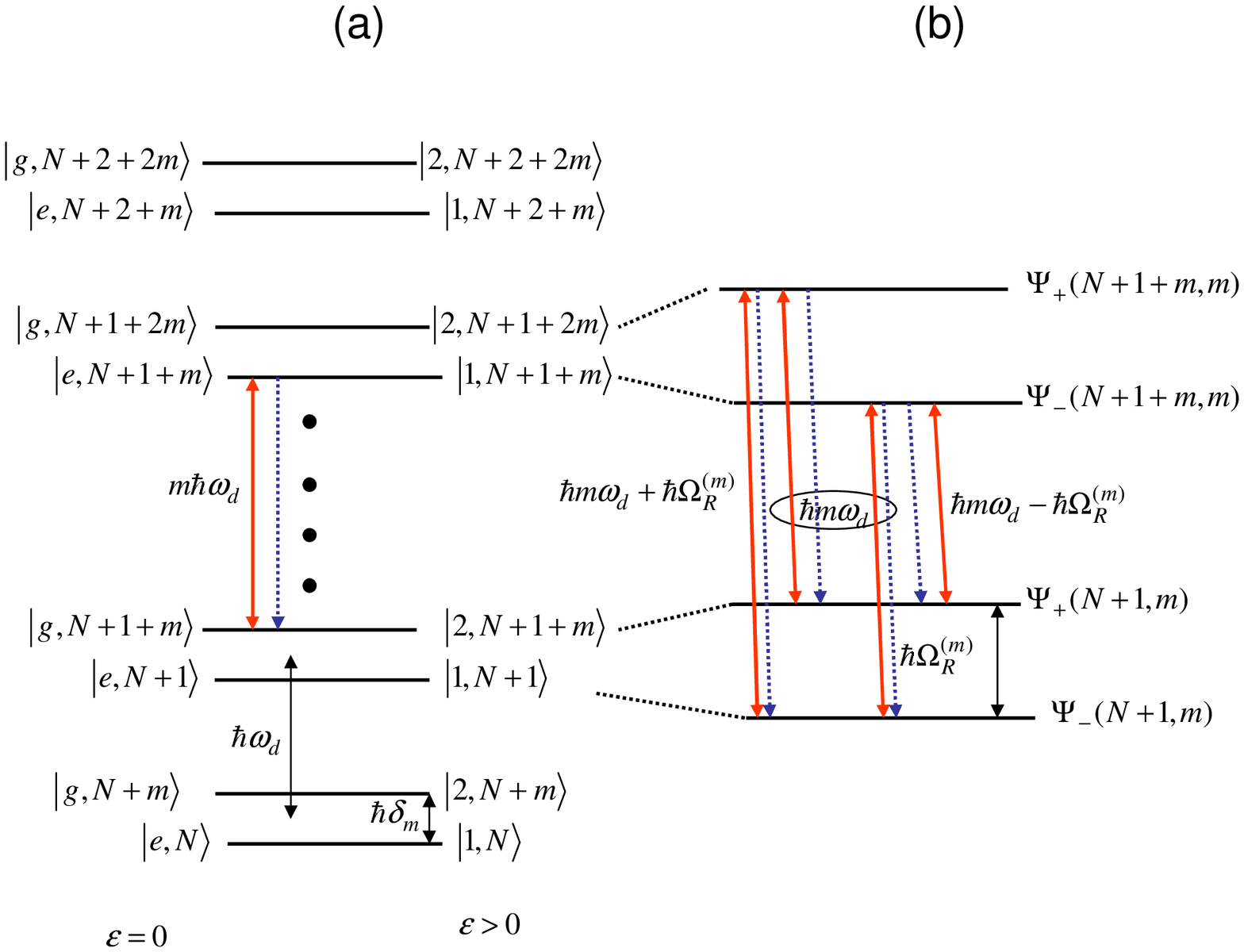}
  \caption{(a) A ladder of the energy doublets which represents the uncoupled
   qubit- photon system ($\varepsilon=0$) and the system where photons are
   coupled to the longitudinal term
   ($\varepsilon>0$). Here $\delta_m=\omega_q-m\omega_d<0$.
Spontaneous and stimulated transitions at $m\omega_d$ are shown by
dotted (blue) and red arrows, respectively. (b) Two doublets of
the singly dressed states of the infinite ladder of energy levels.
Sideband transitions at $\omega_d\pm\Omega_R^{(m)}$ and central
transition at $\omega_d$ are shown by red solid arrows.
Corresponding spontaneous transitions are shown by dotted arrows.
}\label{ladder1}
\end{figure}

We define the uncoupled basis for Hamiltonian (\ref{4}) as a
direct product of the qubit ground $|g\rangle$ and excited
$|e\rangle$ states and Fock states $|N\rangle$ for $N$ driving
photons: $|e,N\rangle$, $|g,N\rangle$. The energy levels for the
uncoupled qubit-photon system are $E_{\pm N}=\hbar\omega_d
N\pm\hbar\omega_q/2$. Having in mind to consider the multiphoton
transitions we assume the frequency of the driving field to be
close to a fraction of the qubit frequency
$\omega_d\approx\omega_q/m$, where $m$ is a positive integer
($m=1, 2,$ etc.) \cite{Ilichev10, Temchenko11}. The structure of
the energy levels for uncoupled qubit- photon system for
$\delta_m=\omega_q-m\omega_d<0$ is shown at the left side of
Fig.\ref{ladder1}a. This structure is a ladder of doublets split
by $\hbar\delta_m$ and separated by $\hbar\omega_d$.

The first three terms of Hamiltonian (\ref{4}) can be reduced to
diagonal form exactly \cite{Coh-Tan}. The eigenenergies and
eigenfunctions are as follows :
\begin{equation}\label{5}
    E_{1N}  = \frac{{\hbar \omega _q }}{2} + \hbar \omega _d N -
\frac{{\hbar g_d^2 \varepsilon ^2 }}{{\omega _q^2 \omega _d }}
\end{equation}
\begin{equation}\label{6}
    E_{2N}  = -\frac{{\hbar \omega _q }}{2} + \hbar \omega _d N -
\frac{{\hbar g_d^2 \varepsilon ^2 }}{{\omega _q^2 \omega _d }}
\end{equation}
\begin{equation}\label{7}
    \left| {1N} \right\rangle  = e^{ - \lambda (d^ +   - d)} \left| {eN} \right\rangle
\end{equation}
\begin{equation}\label{8}
    \left| {2N} \right\rangle  = e^{ \lambda (d^ +   - d)} \left| {gN} \right\rangle
\end{equation}
where $\lambda  = g_d \varepsilon/\omega _q \omega _d $.

Remarkably, the account for the third term in Eq. (\ref{4})
results in essentially the same energy levels as those for the
uncoupled system if we disregard the constant downshift term in
Eqs. (\ref{5}) and (\ref{6}). We show these energy levels at the
right side of Fig.\ref{ladder1}a with the new eigenfunctions
$|1,N\rangle$, $|2,N\rangle$ attributed to them.

The efficient coupling of the driving field with the qubit is
caused by the fourth term in Hamiltonian (\ref{4}). It lifts the
degeneracy of the levels $E_{1N_1}$ and $E_{2N_2}$ with
$\omega_q\approx m\omega_d$, where $m=N_2-N_1$ and gives rise to
the well known dressed energy spectrum shown in Fig.\ref{ladder1}b
(see, for example \cite{Coh-Tan} and references therein).

The calculation of the corresponding matrix element yields the
following result:
\begin{eqnarray}\label{Bessel}
 &  \langle{1, N}|{\sigma _X}\left( {d^+ + d}\right)|2,N+m \rangle
 = & \nonumber \\
&=\exp(-2\lambda^2)\left(\frac{m}
    {\lambda }-4\lambda\right)J_{m} \left( {4\lambda \sqrt {\left\langle N \right\rangle } }
    \right)&
\end{eqnarray}
In Eq. (\ref{Bessel}) $J_m$ is the Bessel function, $<N>$ is the
average number of driving photons which is assumed to be large:
$N_1, N_2\approx\langle N\rangle>>m$.

Hence, for the energy splitting between degenerate states we
obtain:
\[
{E_ \pm }(N,m) = \hbar {\omega _d}N \pm \frac{1}{2}\hbar
\Omega_R^{(m)}
\]

where,
\begin{equation}\label{9}
  \Omega _R^{(m)}  = \sqrt {(\omega _q  - \left| m \right|\omega _d )^2  + 4\Lambda_m
  ^2}
\end{equation}
is the $m$- photon Rabi frequency with
\begin{equation}\label{10}
    \Lambda_m  = g_d \frac{\Delta }{{2\omega _q }}\frac{m}
    {\lambda }J_{m} \left( {4\lambda \sqrt {\left\langle N \right\rangle } } \right)
\end{equation}
which follows from (\ref{Bessel}) for $\lambda<<1$.

This Bessel-like behavior of the Rabi splitting can be obtained
also classically and, in fact, is known for a long time (see, for
example, Ref. \cite{Kmetic86}). It has also been used for the
study of quantum dynamics in the microwave irradiated
superconducting charge qubits \cite{Nakamura01, Krech05, Wilson07,
Wilson10}. However, due to the presence of the photon operator~
$(d^++d)$, the matrix element (\ref{Bessel}) differs from that for
a charge qubit, and from that considered in Ref. \cite{Coh-Tan}.

Similar to the uncoupled spectrum the dressed spectrum consists of
a series of doublets split by $\hbar\Omega_R^{(m)}$ and separated
by $\hbar\omega_d$. As is evident from Eq. (\ref{9}), the
splitting between dressed levels can be effectively adjusted by
three external parameters: the external magnetic flux, $\Phi_X$
which is hidden in $\varepsilon$, the frequency of the driving
field, $\omega_d$, and the input power, which is proportional to
$\langle N\rangle$. For reasons which will be clear later we call
these dressed states as singly dressed states (SDS).

At the degeneracy point $(\epsilon=0)$ only single photon
transition $(m=1)$ survives
\begin{equation}\label{9a}
    \Omega _R^{(1)}  = \sqrt {(\omega _q  - \omega _d )^2
    + \frac{4g_d^2\Delta^2\langle N\rangle}{\omega_q^2}}
\end{equation}

The state vectors for two states of a doublet are superposition
states of the unperturbed states $|1,N_1\rangle$ and
$|2,N_2\rangle$ where $N_2-N_1=m$:

\begin{equation}\label{PsiSDS}
    \Psi _ \pm (N,m) = \alpha _ \pm ^{(m)}\left| {1,{N}}
\right\rangle  + \beta _ \pm ^{(m)}\left| {2,{N+m}} \right\rangle
\end{equation}

where
\begin{equation}\label{alp}
    \alpha _ \pm ^{(m)} =
\frac{{ - \sqrt 2 {\Lambda _m}}}{{\sqrt {\Omega _R^{(m)}(\Omega
_R^{(m)} \mp {\delta _m})} }}
\end{equation}

\begin{equation}\label{bet}
    \beta _ \pm ^{(m)} = \frac{{{\delta _m} \mp \Omega
_R^{(m)}}}{{\sqrt {2\Omega _R^{(m)}(\Omega _R^{(m)} \mp {\delta
_m})} }}
\end{equation}

and ${\delta _m} = {\omega _q} - \left| m \right|{\omega _d}$ is
the detuning between the qubit frequency and the $m$ photon
driving signal frequency.

There are three types of transitions between levels shown in
Fig.\ref{ladder1}: the spontaneous emission, the absorption, and
the stimulated emission caused by an external probe signal. The
allowed transitions occur only between the states with nonzero
matrix element of $\sigma_X$ which is a part of the qubit "dipole"
operator~ $\varepsilon\sigma_Z/\omega_q+\Delta\sigma_X/\omega_q$
which describes the qubit interaction with the photon field
(\ref{H_Int}).

Spontaneous emission is due to the transition between qubit states
with the same number of driving photons $|e,N\rangle\rightarrow
|g,N$ ($|1,N\rangle\rightarrow |2,N$). In Fig.\ref{ladder1}a this
transition is shown by the blue dotted arrow. In the picture of
dressed states this transition corresponds to two sidebands at the
frequencies $m\hbar\omega_d\pm\hbar\Omega_R^{(m)}$ and one central
line $m\hbar\omega_d$ which are shown in Fig.\ref{ladder1}b by the
blue dotted arrows. This is well known fluorescent Mollow triplet
\cite{Mollow72}. Absorption and stimulated emission are shown in
Fig.\ref{ladder1} by the red solid arrows. These transitions occur
between the same levels which are connected by the spontaneous
transitions. Which process is dominant at the given transition,
the absorption or the stimulated emission, or, in other words,
whether the probe signal is amplified or attenuated depends on the
population of the corresponding levels.

By using superconducting qubits as artificial atoms these effects
at sideband Rabi transitions between SDS have  been demonstrated
experimentally. A lasing action with a single Josephson junction
charge qubit has been realized \cite{Astafiev07} and the
stimulated emission by a flux qubit was successfully employed for
the amplification of a microwave signal passing through an open
transmission line \cite{Astafiev10a}. The Rabi sidebands have been
observed for different structures of superconducting qubits
coupled to a microwave transmission line: one photon sidebands for
three level dressed states in a transmon qubit \cite{Koshino13},
two-photon sidebands for a Cooper pair box \cite{Wallraff07} and
multiphoton sidebands for a flux qubit \cite{Shimazu13}.

Therefore, in principle, superconducting qubits allow one to
obtain the amplification (attenuation) of a probe signal at the
Rabi sideband transitions, which are similar to those in quantum
optics.

\subsection{Interaction of the driven flux qubit with a quantized
photon field in a coplanar waveguide resonator}

In this section we consider the effect of amplification
(attenuation) of a probe microwave signal which is due not to the
sideband Rabi transitions, but to the transitions directly at the
Rabi frequency. In spite of some analogy with quantum optics,
there is an essential difference: in quantum optics direct
transitions at the Rabi frequency are forbidden at least in the
first order perturbation since the matrix element of the dipole
interaction between Rabi levels is exactly equal to zero. As
opposed to optical systems, solid state quantum systems are
scalable and tunable, therefore the effect of direct Rabi
transitions can be used for microwave quantum engineering
\cite{Saiko10}.

We place the system qubit+ driving field (\ref{4}) in a photon
resonator (low loss rate coplanar waveguide) with the frequency
$\omega_0$ being close to Rabi frequency $\Omega_R^{(m)}$. This
results in the total Hamiltonian, $H=H_{qb-d}+H_r$, where

\begin{equation}\label{15}
    H_r = \hbar \omega _0 a^ +
a + \hbar g_{_r } \frac{\varepsilon }{{\omega _q }}\sigma _Z
\left( {a^ +   + a} \right) + \hbar g_{_r } \frac{\Delta }{{\omega
_q }}\sigma _X \left( {a^ +   + a} \right)
\end{equation}
Here $g_r$ is the qubit resonator coupling; $a^+, a$ are creation
and annihilation operators for photons in the resonator, and we
assume that the number $n$ of resonator photons is small ($n<<N$).

Since the frequency of the resonator $\omega_0$ is assumed to be
smaller than the qubit gap~ $\Delta$, the resonator cannot excite
the qubit. Therefore we neglect the $\sigma_X$ term in Eq.
(\ref{15}). The second term in Eq. (\ref{15}) evidently, leads to
the transition between two levels within a dressed doublet
(Fig.\ref{ladder1}b), a transition which is not allowed for
$\sigma_X$. It means that a probe signal which is matched to the
splitting between the dressed states of the same doublet can cause
resonant transitions between them. This leads to the emission or
absorption of Rabi photons depending on the populations of these
nearby levels resulting in the energy exchange between the flux
qubit and the resonator. Below we analyze this resonant effect
within the frame of doubly dressed states (DDS).

\begin{figure}
  \includegraphics[height=.4\textheight, angle=-90]{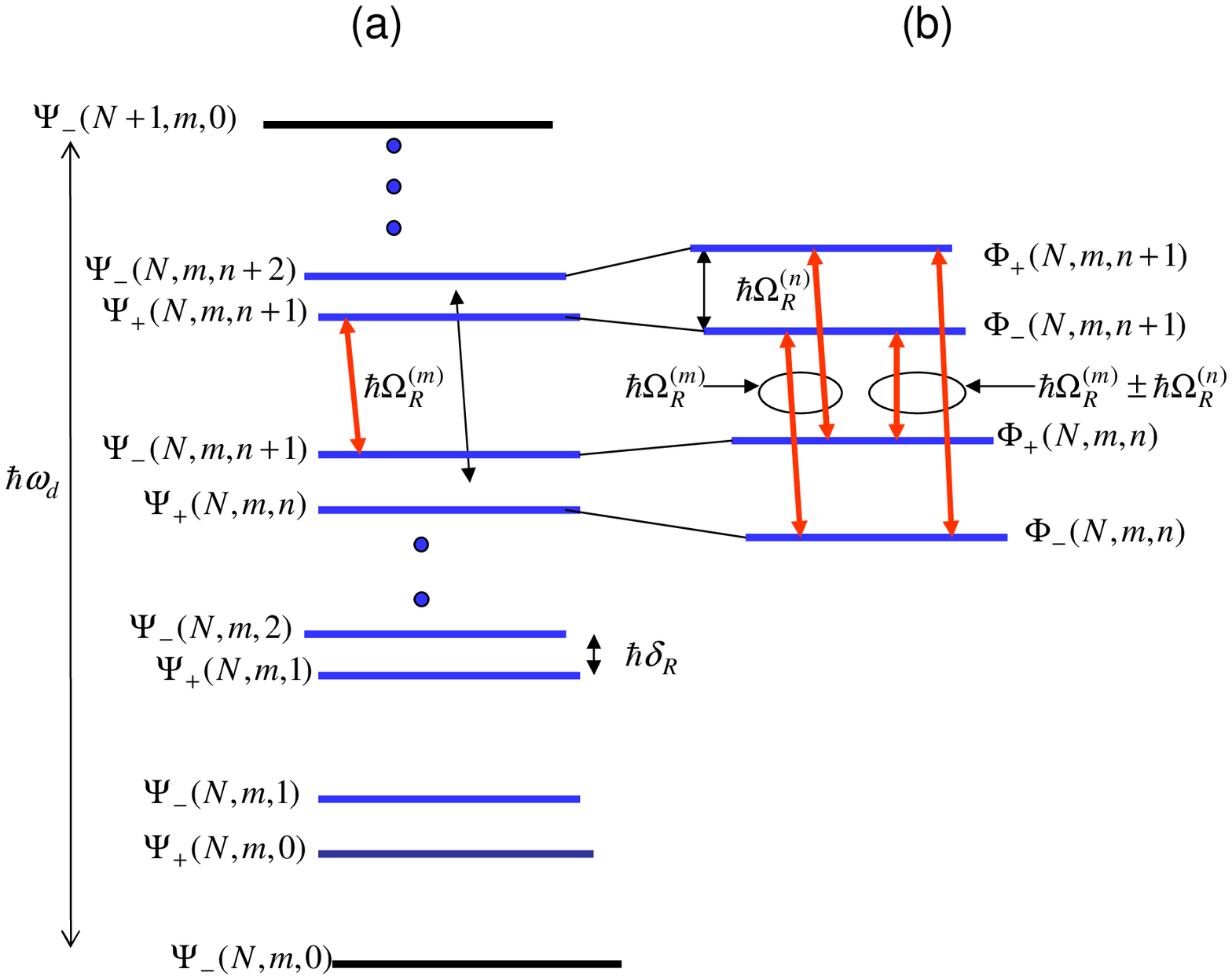}
  \caption{(a) Doublets of the uncoupled system of singly dressed states and $n$- photon
resonator ($\delta_R=\Omega_R^{(m)}-\omega_0<0$). The allowed
transition induced by the longitudinal term $\sigma_Z$ is shown by
the red arrow. (b) Two pairs of the doubly dressed states. Two
sidebands at $\Omega_R^{(m)}\pm\Omega_R^{(n)}$ and a central line
at $\Omega_R^{(m)}$  are shown by red arrows.}\label{ladderDDS}
\end{figure}

We define, analogous to Eqs. (\ref{7}) and (\ref{8}), the wave
functions $\left| {1Nn} \right\rangle  = e^{ - \lambda (d^ +   -
d)} \left| {eNn} \right\rangle$ and $\left| {2Nn} \right\rangle  =
e^{ \lambda (d^ +   - d)} \left| {gNn} \right\rangle$ with the
eigenenergies $E_{1N}=\hbar\omega_q/2+\hbar\omega_d
N+\hbar\omega_0 n$, $E_{2N}=-\hbar\omega_q/2+\hbar\omega_d
N+\hbar\omega_0 n$,  where $n$ is the number of Rabi photons in
the resonator. Furthermore, the wave functions
$\Psi_{\pm}(N,m)\bigotimes |n\rangle$ with $n$ photons we denote
as $\Psi_{\pm}(N,m,n)$. If the resonator frequency $\omega_0$ is
close to the Rabi frequency $\Omega_R^{(m)}$ we get an energy
spectrum a part of which is shown in Fig.\ref{ladderDDS}a. These
doublets fill the gap between two neighbor doublets of the dressed
states.

The additional degeneracy for the levels $\Psi_-(N,m,n+1)$ and
$\Psi_+(N,m,n)$ is lifted by the second term in the resonator
Hamiltonian (\ref{15}). For the energy splitting between these new
degenerate states we find:
\[
{E_ \pm }(N,m,n) = \hbar {\omega _d}N +\hbar\omega_0 n\pm
\frac{1}{2}\hbar \Omega _R^{(n)}
\]
where
\begin{equation}\label{DDSen}
    \Omega _R^{(n)} = \sqrt {{{\left( {\Omega _R^{(m)} - {\omega _0}} \right)}^2}
     + 4{F_n^2}}
\end{equation}
with
\begin{equation}\label{DDF_n}
 F_n = \frac{{2{\Lambda _m}}}{{\Omega
_R^{(m)}}}\frac{{g_r\varepsilon }}{{{\omega _q}}}\sqrt {n + 1}
\end{equation}

The resulting wave functions are
\begin{equation}\label{dds}
    \Phi_{\pm}(N,m,n)=A_{\pm}\Psi_+(N,m,n)+B_{\pm}\Psi_-(N,m,n+1)
\end{equation}
where
\[
\Psi _ \pm (N,m,n) = \alpha _ \pm ^{(m)}\left| {1,{N},n}
\right\rangle  + \beta _ \pm ^{(m)}\left| {2,{N+m},n}
\right\rangle
\]
with $\alpha_{\pm}^{(m)}$, $\beta_{\pm}^{(m)}$ being given in Eqs.
(\ref{alp}), (\ref{bet}). The quantities $A_{\pm}$, $B_{\pm}$ in
Eq. (\ref{dds}) are as follows:
\begin{equation}\label{A}
    A_ \pm ^{} = \frac{{ - \sqrt 2 F}}{{\sqrt {\Omega _R^{(n)}(\Omega _R^{(n)}
    \mp {\delta _R})} }}
\end{equation}
\begin{equation}\label{B}
    B_ \pm ^{} = \frac{{{\delta _R} \mp \Omega _R^{(n)}}}
    {{\sqrt {2\Omega _R^{(n)}(\Omega _R^{(n)} \mp {\delta _R})} }}
\end{equation}
where $\delta_R=\Omega_R^{(m)}-\omega_0$.

Two pairs of states (\ref{dds}) are shown in Fig.\ref{ladderDDS}b.
These states are called the doubly dressed states since they are
obtained by the dressing of the previously dressed states
(Fig.\ref{ladder1}b) with the interaction between the qubit and
the photons in in the resonator.

The longitudinal term in the qubit dipole operator, $\sigma_Z$, of
the probe signal induces the transitions between singly dressed
states $\Psi_+(N,m,n)$ and $\Psi_-(N,m,n)$. An example of such a
transition is shown by the red arrow in Fig.\ref{ladderDDS}a. In
the picture of doubly dressed states (Fig.\ref{ladderDDS}b) this
transition corresponds to two sideband lines with the frequencies
$\Omega_R^{(m)}\pm\Omega_R^{(n)}$, and a central line
$\Omega_R^{(m)}$. If $\omega_0=\Omega_R^{(m)}$ we get the only
line  at $\Omega_R^{(n)}$.~ Hence, the transition between the
states of singly dressed doublet transforms to sideband
transitions between neighbor doublets of doubly dressed states.

\section{Transmission factor. Amplification and attenuation of a probe signal}

There are several external parameters which can be used to match
the splitting $\Omega_R^{(m)}$ between Rabi levels with the
fundamental frequency $\omega_0$. These are the external magnetic
flux $\Phi_X$ (\ref{2}), the driving frequency $\omega_d$, and the
input driving amplitude $A_d=g_d\sqrt{\langle N\rangle}$.

While it is not difficult to find the positions of the resonance
from the equation $\omega_0=\Omega_R^{(m)}$, it is not easy to
calculate the transmission factor which shows the amplification or
attenuation of the probe signal. The reason for this is that the
dynamics of the qubit is essentially dissipative. The bare qubit's
relaxation, $\Gamma$ and decoherence, $\Gamma_{\varphi}$ rates are
transformed in a complicated way under influence of the
interaction of the qubit with photon fields \cite{Hauss08b,
Shevchenko14}. These effects of dissipation can be accounted for
by the Liouville equation for the density matrix of the system,
including the relevant damping terms, which are assumed to be of
Markovian form \cite{Shevchenko14}.

The transmission factor, which is the output-to-input ratio for a
probe signal, is defined in the way adopted in quantum optics
\cite{Shevchenko14}:

\begin{equation}\label{tf}
    t=i\frac{\hbar\kappa}{2\xi_p}\langle a\rangle,
\end{equation}

where $\kappa$ is the loss rate of photons in the coplanar
waveguide, $\xi_p$ the amplitude of the probe signal, $\langle
a\rangle=Tr(a\rho)$, $\rho$ the density matrix of the system, and
the trace over the photon states of the fundamental mode
$|n\rangle$ and the two qubit states $|\pm\rangle$.

\begin{figure*}[ht]
  \includegraphics[width=15 cm]{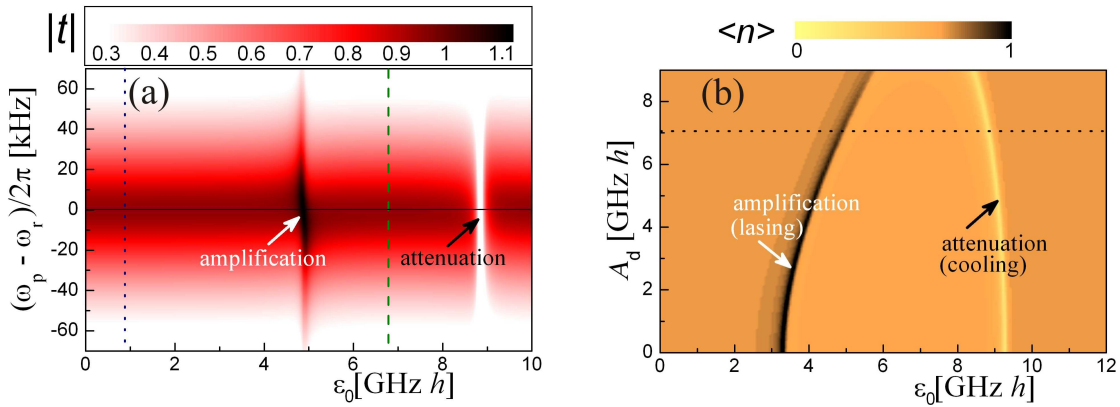}
\caption{(left) Normalized transmission amplitude $|t|$ for the
fundamental- mode signal versus the bias $\varepsilon$ and the
probing frequency detuning $\omega_p-\omega_0$. The amplification
and attenuation of the transmitted signal is displayed as the
increase or decrease of the transmission at resonance
$\omega_p\approx\omega_0$; (right) The average intracavity photon
number $\langle n\rangle$
  versus the bias $\varepsilon$
and driving amplitude $A_d$. The value of the driving amplitude,
at which left plot of the Fig.\ref{tn} was calculated, is shown by
the dashed line. The dark and light colors correspond to the
increase and decrease of the intracavity photon number which, can
be termed as lasing or cooling of the resonator and results in the
amplification or attenuation of the transmitted probing signal.
Here $\omega_p=\omega_0$. }\label{tn}
\end{figure*}

Here we show two plots which have been calculated with the account
of dissipation effects \cite{Shevchenko14}. In Fig.\ref{tn}(left)
we show the transmission amplitude $|t|$ of the fundamental-mode
signal as a function of the qubit's bias $\varepsilon$ and the
detuning $\omega_p-\omega_0$ from the probing frequency
$\omega_p$. When the frequency of the probing signal is close to
the Rabi frequency $\Omega_R^{(m)}$ a resonant energy exchange
between the qubit and the fundamental mode of the resonator
results in amplification or attenuation of the transmitted signal.
The observation of such amplification was recently reported in
\cite{Oelsner13}.

The effect of amplification and attenuation of the transmitted
signal can also be related to the increase or decrease of the
intracavity photon number. This is demonstrated in
Fig.\ref{tn}(right), where the stationary solution for the
intracavity photon number is plotted as the function of the
qubit's bias $\varepsilon$ and the driving amplitude $A_d$.

These calculations were done in the regime of the weak driving for
the following parameters \cite{Oelsner13}: $\Delta$ = 3.7 GHz,
$g_r$ = 0.8 MHz, $\omega_0/2\pi$= 2.5 GHz, $\Gamma$ = 80 MHz,
$\Gamma_{\varphi}$ = 10 MHz,  $A_d$ = 7 GHz, $\omega_d=3\omega_0$,
~$\kappa$= 30 kHz, and $\xi_p=0.5\kappa$.

\begin{figure*}[ht]
\includegraphics{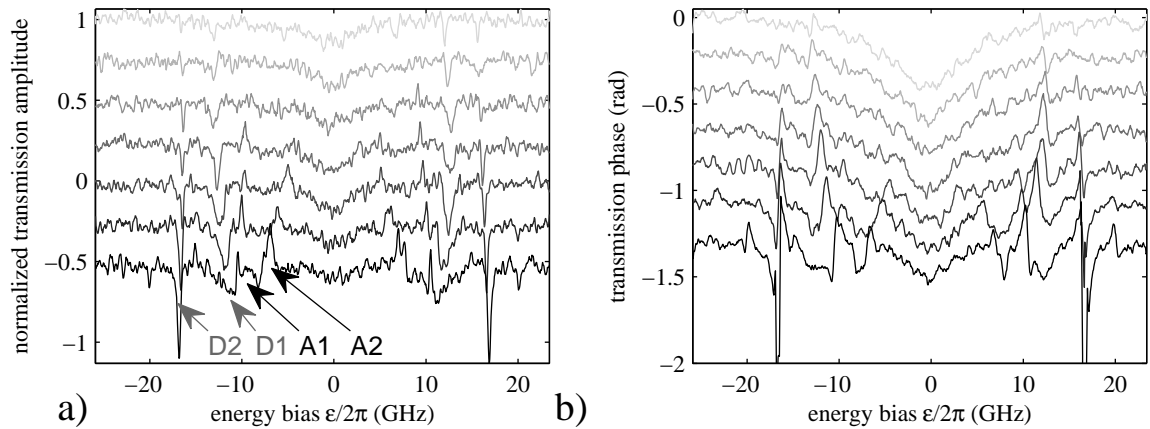}
  \caption{Normalized transmission amplitude (a) and phase (b) of a
  weak probe signal at $\omega_p = \omega_0$ through the resonator.
  From light to dark color the driving power is increased in 2~dBm steps.
  The curves are shifted from another by 0.25 for the amplitude and 0.2 rad for
  the phase for better visibility. In (a) the peaks for the one (A1) and two
  (A2) photon amplification and similar the dips of the attenuation (D1 and D2)
  are marked. }\label{G1}
\end{figure*}

Below we show the results of a recent experiment where a
substantial amplification of a probe signal is demonstrated. We
used a sample similar to the one in \cite{Oelsner10}. The
resonator is probed at its fundamental mode with a frequency of
$\omega_0/2\pi \approx$ 2.5~GHz and driven at its 5th harmonic
with $\omega_d = 5 \omega_0$. If the transmission is monitored
with low driving powers, only the dispersive shift of the
resonator frequency close to the qubits degeneracy point is
observed. No resonant interaction occur since the energy gap of
the qubit $\Delta \approx$ 6~GHz is bigger than the eigenfrequency
of the resonators fundamental mode. This can be seen in
Fig.~\ref{G1} at the upper, light curve, which is measured for low
driving powers.

\begin{figure*}[ht]
\includegraphics{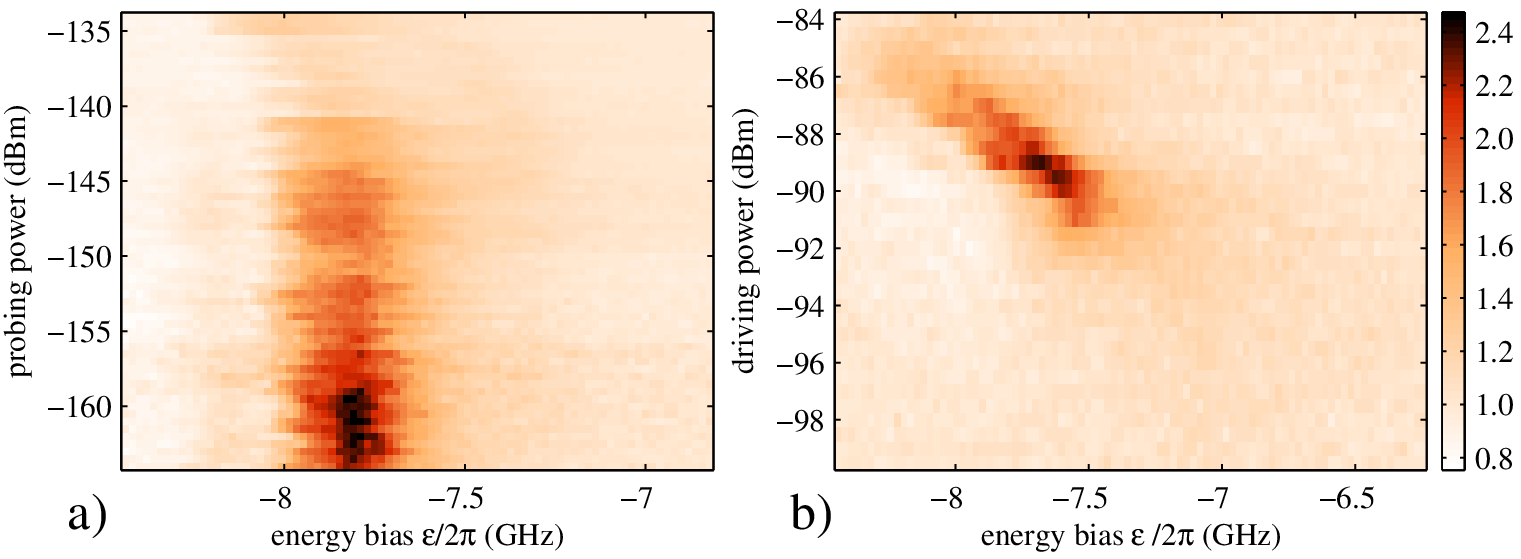}
  \caption{Dependence of the amplification on the probing (a) and driving (b)power at the
  input of the resonator measured at $\omega_0$. The colormap for both ofthe plots is the
  same and the corresponding normalized transmission is given in the left colorbar.The left
  plot is measured with a driving power of -90~dBm and the right plot with a probingpower
  of -160~dBm. }\label{G2}
\end{figure*}

When the driving power is increased (to the darker and lower
curves in Fig.~\ref{G1}), signatures of resonant interactions
appear. They correspond to Rabi resonances between the dressed
states and the resonator fundamental mode and manifest themselves
as dips and peaks in Fig.~\ref{G1} (a) and as characteristic phase
jumps in Fig.~\ref{G1} (b). The latter is expected for avoided
crossings of the energy levels of the resonator (compare for
example \cite{Oelsner10}). The amplification peaks occur, where
the qubit energy is smaller than the driving frequency, where
$\left| \epsilon\right| <$ 11~GHz, and vice versa for the
attenuation. This is because the population depends on the sign of
the detuning between the two, as explained in detail in
\cite{Oelsner13,Shevchenko14}. The shift of the resonance points
follows the resonance condition $\Omega_R^{(1)} = \omega_0$. Note,
if the amplitude term in (\ref{9a}) dominates, the detuning is
close to zero to fulfill the resonance condition. Then the
populations get equalized and a wide dip instead of a
peak-dip-structure is expected due to the frequency shift in
resonance. Finally we mention that not only the position of the
peaks, but also their height depends on the driving power. To find
optimal parameters for the amplification we analyzed the power
dependencies for one of the amplification points as presented in
Fig.~\ref{G2}.

As seen in Fig.~\ref{G2} (a) the probing power has, in the
analyzed region, almost no influence on the position of the
resonance point. Only for high powers above -140~dBm a small shift
to bigger values of $\left| \epsilon \right|$ is observed.
Nevertheless, as also the amplification effect vanishes in this
region no further investigations for higher powers where done. The
gain of the feature has a monotonic increase for decreasing
probing powers. This may be misleading since the resonant process
can add a constant number of photons in each cycle of the
resonator. Only the reference signal is reduced and since the
transmission is given as the ratio of the output to the input
amplitude the described dependence is well expected. The driving
power in turn has influence on both the position and the strength
of the amplification, as seen in Fig.~\ref{G2} (b). As already
briefly discussed the position follows the resonance condition.
Therefore, if the amplitude dependent term of the Rabi splitting
is increased the detuning between the driving signal and the qubit
needs to be reduced. Also a clear optimum in the amplification
process is observed. This has two reason, on the one hand the
effective coupling between the Rabi levels and the resonator is
increased with increasing driving amplitude (see (\ref{DDF_n}) or
\cite{Oelsner13, Shevchenko14}), but also as explained above the
population difference between the states is decreased (see also
the dissipative dynamics in \cite{Hauss08a, Shevchenko14}). In
addition to the increasing coupling also the eigenfrequency of the
resonator at the Rabi frequency is shifted more strong. Then by
measuring only for $\omega_0/2\pi $= const = 2.5~GHz we not always
measure in the maximum of the resonators Lorentzian, which can
hide the amplification.

\section{Conclusion}
We have considered a superconducting flux qubit placed in a
microwave cavity resonator. The qubit is irradiated by a strong
driving field at a higher harmonic and is probed by a weak signal
at its fundamental frequency (first harmonic). The driving signal
is not  observed directly and is included into the considerations
by the qubit's dressed states. Similarly, the interaction of the
dressed qubit with the resonator's fundamental mode can be
described as the second dressing of the qubit's dressed states.
When the energy of the probing photons matches the dressed energy
levels, a resonant energy exchange between the qubit and the
resonance cavity results in either the amplification or the
attenuation of the probing signal, depending on the mutual
populations of Rabi levels, which, in turn, depend on the
frequency detuning and on the tunable relaxation rates.

We briefly reviewed the major theoretical and experimental
activity in this field and presented recent calculations and
experimental results which demonstrate experimentally the main
features of the dressed-state amplification of a probe field by a
strongly driven qubit directly at the Rabi frequency. We believe
that further investigation of this effect could find potential
applications in the development of new types of quantum microwave
amplifiers and sources.

\begin{acknowledgments}
This research has received funding from the European Community's
Seventh Framework Programme (FP7/2007-2013) under grant agreement
No. 270843. The work was partly supported by NAS of Ukraine
(Project No. 4/14 -NANO), DKNII (Project No. M/231-2013), BMBF
(UKR-2012-028). ~Ya.S.G. acknowledges the financial support from
the Russian Ministry of Education and Science within the framework
of the project of the state assignment, entitled "Development of
the basic components for quantum microwave circuits".
\end{acknowledgments}

\end{document}